\def\sii{S~{\sc ii}}

\def\hi{H~{\sc i}}

\def\feii{Fe~{\sc ii}}

\def\siii{Si~{\sc ii}}

\def\znii{Zn~{\sc ii}}

\documentclass[useAMS,usenatbib,epsfig]{mn2e}

\usepackage{rotating}
\usepackage{float}
\usepackage{natbib}
\usepackage{verbatim}
\usepackage{hyperref,graphicx}

\title[Dust depletions and extinctions]{Metals, depletion and dimming:  decrypting dust}
\author[Zafar \& M{\o}ller] {Tayyaba Zafar$^{1}$\thanks{e-mail:tayyaba.zafar@mq.edu.au} and Palle M{\o}ller$^2$ \\
$^1$ Australian Astronomical Optics, Macquarie University, 105 Delhi Road, North Ryde, NSW 2113, Australia. \\
$^2$ European Southern Observatory, Karl-Schwarzschild-Strasse 2, 85748, Garching, Germany. \\
}

\begin{document}


\pagerange{\pageref{firstpage}--\pageref{lastpage}} \pubyear{2018}

\maketitle

\label{firstpage}

\begin{abstract}
Dust plays a pivotal role in the chemical enrichment of the interstellar medium. In the era of mid/high resolution spectra and multi-band spectral energy distributions, testing extinctions against gas and dust-phase properties is becoming possible. In order to test relations between metals, dust and depletions, and comparing those to the Local Group (LG) relations, we build a sample of 93 $\gamma$-ray bursts and quasar absorbers (the largest sample so far) which have extinction and elemental column density measurements available. We find that extinctions and total column density of the volatile elements (Zn, S) are correlated (with a best-fit of dust-to-metals (DTM) $4.05\times10^{-22}$\,mag\,cm$^2$) and consistent with the LG DTM relation. The refractory elements (Fe, Si) follow a similar, but less significant, relation offset about 1\,dex from the LG relation. On the assumption that depletion onto dust grains is the cause, we compute the total (gas$+$dust-phase) column density and find a remarkable agreement with the LG DTM relation: a best-fit of $4.91\times10^{-22}$\,mag\,cm$^2$. We then use our results to compute the amount of `intervening metal from unknown sources' in random sightlines out to redshifts of $z=5$. Those metals implicate the presence of dust and give rise to an average `cosmic dust dimming' effect which we express as a function of redshift, CDD($z$). The CDD is unimportant out to redshifts of about 3, but because it is cumulative it becomes significant at redshifts $z=3-5$. Our results in this paper are based on a minimum of assumptions and effectively relying on observations.
\end{abstract}

\section{Introduction}
Interstellar dust plays a crucial role in the chemical enrichment of the interstellar medium (ISM) and is strongly linked with star formation \citep[e.g.][]{cortese12}. Extinction, $A_V$, is the scattering and absorption of photons along the travel path from a source to the observer. Depletions of heavy elements in the ISM are another way to study dust properties due to its association with dust \citep{jenkins87}. Depletion is a strong indicator that gas phase refractory element ejected from stars is efficiently condensed onto ISM dust grains. The dust grains are primarily made up of O, Si, C, Mg, and Fe metals \citep{draine03,decia16} which are introduced into the ISM through the stellar winds during the stellar evolution or at the end of the life of a star. However, there is still a debate over the origin of the bulk of dust, where various candidates have been proposed such as: $i)$ dense envelopes of evolved, low-mass asymptotic giant branch (AGB) stars \citep{gail09}, $ii)$ condensation in supernova ejecta \citep{dunne03}, and $iii)$ grain growth in the dense molecular clouds \citep{draine09}. 

The dust-to-metal ratios over cosmic time provide information on the interplay between dust and gas in the ISM of galaxies. Theoretical studies show dust-to-metals ratios to remain constant over time and metallicity, provided both dust and metals are produced in and ejected from stars \citep[e.g.,][]{franco86}. In a sample of gamma-ray burst (GRB) afterglows, foreground quasar absorbers, and lensed galaxies \citep{chen13}, \citet{zafar13} found that the dust-to-metals ratios remain constant over a wide range of redshifts, metallicities, hydrogen column densities, and galaxy types. The dust-to-metals ratios in their sample are consistent with the Local Group (LG) relation. A dust model for cosmological simulations shows that for a constant dust-to-metals ratios most of the dust is produced in supernovae \citep{mattsson14,mckinnon16}. There is a discrepancy between using direct line-of-sight extinctions and depletion derived dust measures. The depletion derived dust content suggests dust-to-metals are correlated with metallicity  \citep{decia16,wiseman17,devis17}. Using depletion derived dust methods, individual objects \citep{watson06,savaglio12,friis15} and samples \citep{wiseman17} find mis-match with the observed $A_V$. For such a scenario models \citep{mattsson14,schneider16,zhukovska16,popping17} suggest grain growth in the ISM is dominant dust production pathway.

Extinction is caused by the dust particles in the ISM, therefore, $A_V$ should scale with the column density of atoms in the dust-phase. The Milky Way (MW) refractory elements dust-phase column density and $A_V$ follow each other \citep{vladilo06}. A small sample of damped Ly$\alpha$ absorbers (DLAs) along the line of sights of quasars (QSOs) are reported to follow the MW relation \citep{vladilo06}. A similar correlation is also seen for a small sample of GRBs where exclusion of the cases with a 2175\,\AA\ bump feature strengthen the correlation (\citealt{decia13}; see however \citealt{wiseman17} but with a large scatter). 

Dust-to-metal ratios for DLAs have also been derived from depletions using depletion corrections from the Galactic or disk$+$halo environments \citep{savaglio01,decia13}. In this paper, we test the connections between extinction and depletions in a significantly larger sample combined of both GRBs and QSO-DLAs, but
here adopting methods with a minimum of assumptions and using mostly direct observables. The aim of our investigation is both to test the above-mentioned relations and test for connections between the two quantities and other observables which lead us to infer effect of dust at cosmological distances.
For this, we need high redshift measurements of extinctions, elemental abundances of refractory and volatile elements, and hydrogen column densities of the systems if possible. In \S2 we present our sample and sample selection criteria. In \S3 we describe our methods used for the investigation together with the results of the analysis. The discussion is provided in \S4 and conclusions are given in \S5. Throughout the paper, errors are 1$\sigma$ unless stated otherwise.

\section{Sample Selection}
We searched the literature carefully and selected all published 
GRB-DLAs and QSO-DLAs sightlines conforming to our requirements
which are as follows. The object must have spectral energy distributions (SEDs) and optical spectroscopic
data available with measurements of $A_V$, column densities of \znii\
and \feii , or of \sii\ and \siii . The GRBs are selected
only if they had their optical extinction derived from simultaneous SED
fitting to X-ray$-$to$-$optical/NIR data using either a single or broken
power-law (see \citealt{zafar11,greiner11,schady12,covino13,bolmer17} for discussion on $A_V$ determination).  
This is a reliable method to determine extinctions at higher redshifts
where the intrinsic slopes are constrained by the X-ray data. Note that there is some degeneracy between broken power-law break frequency and extinction, which could lead to inference of grey dust for some instances \citep{watson06,perley08,friis15}. However, overall a fixed spectral break change ($\Delta\beta=0.5$) between the optical and X-ray slopes is preferred for GRBs \citep{greiner11,zafar11,japelj15}. For QSO-DLAs, reddening must be determined either from QSO colors or extinctions through
template fitting to the QSO SED. Those methods are less robust than the X-ray
supported GRB fits, but are widely adapted. We refer the reader to \citet{zafar15,krogager15,krogager16} for more discussions on $A_V$ determination for QSO. 
The requirement for the pairs of elements (\znii\ and \feii , or of \sii\
and \siii ) are in order to be able to derive depletions. Here Zn and S are
volatile elements and Fe and Si are refractory elements
\citep[e.g.,][]{ledoux02,draine03,vladilo11,decia16}. Defined this way our
initial sample consists of 28 GRBs and 32 QSO-DLAs with the required
measurements available. 
To this we add sources where only part of the required measurements are
complete but where limits have been determined for the rest.
The vast majority of the elemental abundances have been determined via
detailed spectral line fitting, for eight GRBs elemental abundances (or
limits) were derived from rest-frame equivalent widths of non-saturated
lines provided by \citet{fynbo09} as described in \citet{laskar11} and \citet{zafar13} (see references to Table \ref{GRB}).


In total this makes up a sample of 46 GRBs (see Table \ref{GRB}) and 47
QSO-DLAs (see Table \ref{QSO}), i.e. a total of 93 independent sightlines.
The sample is a complete literature sample and as such inherits whichever
biases went into the initial target selection for observation and
publication, but no additional biases were introduced by us other than
the selection definitions we provide above. The sample
covers redshifts from the nearby Universe ($z\approx0.4$)
out to the epoch of reionisation ($z\approx$6.3) and is significantly
larger than any other sample used for previous similar studies.  
Following the discussion of \citet{watson11}, we use solar abundances from \citet{anders89} for metallicity and depletion determinations. \citet{anders89} abundances are better estimates of the typical Galactic ISM \citep[see also][]{asplund09}.

\begin{figure}
\begin{center}
{\includegraphics[width=\columnwidth,clip=]{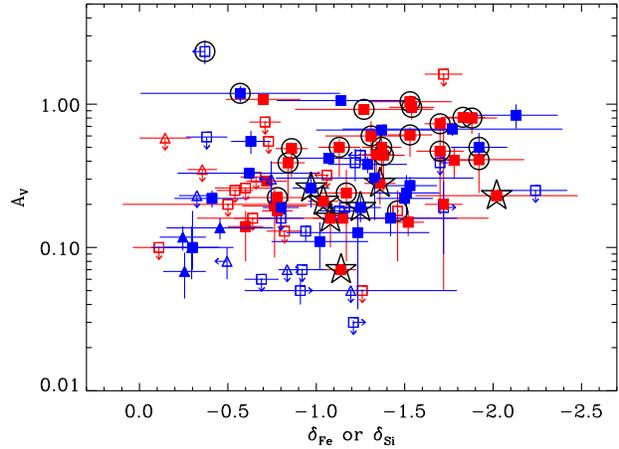}}
\caption{Extinctions against the iron or silicon depletions. The red color corresponds to QSO-DLA cases, while blue indicates GRB-DLA cases. For both, squares represent depletions determined from the [Fe/Zn] depletions and triangles show the [Si/S] depletions. Filled symbols are measurements while open symbols are limits. The black encircled data points highlight cases with a 2175\,\AA\ bump and starred are the cases with N(H$_2$) measurements.}
\label{depletions}
\end{center}
\end{figure}

\section{Methods and results}

The objective of this study is to test which relations are seen in the
complete literature sample we have compiled, partly to clarify some
concepts related to dust, metals, and depletion, but in particular to
help future studies related to, or depending on, dust-to-metal ratios.
We here use the term `dust-to-metal' (or in short DTM) in the same
way as it was introduced in \citet{watson11}.
It is thought that depletion of metals, i.e. the metals `missing' in
sightlines based on a comparison to some expectation, is related to
dust production, and therefore at some level should correlate with
dust absorption. On the other hand, it is not the depletion alone
which defines the amount of dust, obviously the amount of dust will
grow with both the fraction of `missing' metals, but also with the
column density of the absorber. A first test should therefore be to
ask which of the two is a most important quantity.

We therefore first compare the extinctions directly to depletions
$\delta_{\rm (Fe)}$ and $\delta_{\rm (Si)}$ (Fig. \ref{depletions}).
The depletions are calculated by comparing elements 
having a similar nucleosynthetic history. The iron peak element Fe
is compared with iron peak element Zn and likewise $\alpha$-element
compared with the other one from the chain (Si, S) such that
$\delta_{\rm Fe}=$ log (N(\feii)/N(\znii)) $-$ log (Fe/Zn)$_\odot$
and similar for silicon. In Fig. \ref{depletions} we plot
$\delta_{\rm (Fe)}$ for each target if available (squares), and
plot only $\delta_{\rm (Si)}$ (triangles) if $\delta_{\rm (Fe)}$
is not known. As can be seen from the figure there is no obvious
strong correlation in this sample between reddening and depletion.
This visual impression is confirmed by a statistical analysis
which gives a correlation coefficient of $r=-0.16$
($r=-0.05$ ignoring the data providing only limits) with a significance
of only $\approx87\%$. The weak correlation is not very surprising
because of the wide range of \hi\ column densities of the sample.

[Fe/Zn] is found to be a reliable tracer of dust in the ISM \citep{decia16,decia18}. However, for high-redshift sources this measurement is missing because of the available spectral coverage. We, therefore, chose [Si/S] as tracer of dust in high-redshift sources. We also attempted to correct [Si/S] using \citet{decia16} correlations but as we later cannot make better use of it (see \S\ref{metdep}), therefore, we kept using [Si/S] for the cases where [Fe/Zn] is not available.

\subsubsection{Zn and S depletion}
We considered Zn and S as volatile elements to derive equivalent metal column densities. For QSO-DLAs, Zn \citep{decia16} and for Galactic sightlines, S \citep{jenkins09,jenkins17} are reported to be depleted. However, for a sample of 293 QSO-DLAs, \citet{vladilo11} discussed about using S and Zn as volatile elements. We refer to the reader to their section 2.5 for more discussion on selection of these elements. Briefly outlining, S and Zn are depleted in highly depleted Galactic sightlines \citep{jenkins09,jenkins17}. \citet{scappini03} have discussed chemical pathways leading to S depletion in molecular clouds. DLAs are typically less dusty (compared to Galactic sightlines) and have less molecular detection, so Zn and S will be mostly undepleted. \citet{noterdaeme08} and \citet{savage96} find that molecular fraction is correlated with depletions for DLAs and Galactic ISM, respectively. Considering this, we expect Zn or S depletion for the cases where we have high molecular fraction. However, searching in the literature we find only seven (out of 93) cases have molecules detection. Moreover, S is a troublesome element \citep{jenkins17} for Galactic sightlines and for DLAs because of blending with the Ly$\alpha$ forest. Super-solar Galactic S abundances \citep{jenkins09} are found for low column density (N(\hi)$<10^{19.5}$ cm$^{-2}$) sightlines. In case of DLAs, gas is predominantly neutral above N(\hi)$>10^{19.5}$ cm$^{-2}$ (e.g., \citealt{meiring09}, see also \citealt{wolfe05}). In our sample, we mostly have systems with N(\hi)$>10^{19.5}$ cm$^{-2}$ except one, therefore, we should not see such a behaviour.

\subsection{Correlation with metal column densities}\label{metcol}
Total column densities of a given element are direct observables, and
we now wish to test if those correlate with $A_V$ in our sample. In
order to make the data simple to compare to the Local Group $N_{H}-A_V$, and also
to be able to inter-compare different elements, we follow the strategy
used by \citet{zafar13} and first apply a simple modification to the
data. We refer the reader to \citet{zafar13} for more details on the method. For a given object, and a given volatile element `X', we
first determine the metallicity from that element [X/H] and then we
multiply logarithmic metallicity with N(\hi) such that it becomes N(\hi)$\times10^{\rm[X/H]}$. This has the
advantage of shifting each element to the $N_{H_{X}}$ (equivalent
soft X-ray \hi\ column density) used by \citet{watson11} to
determine the $A_V$ of the Milky Way (MW).

\begin{figure}
\begin{center}
{\includegraphics[width=\columnwidth,height=2.8in,clip=]
{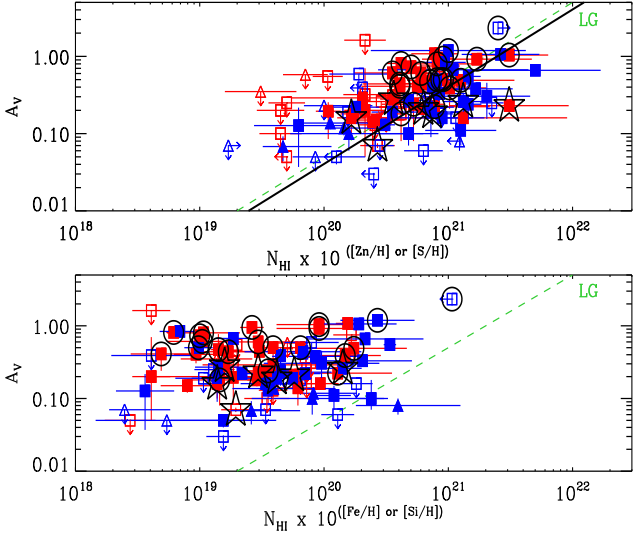}}
\caption{Extinctions against total equivalent column densities. The green dashed line represents the
dust-to-metals ratio for the LG environments with the green dotted lines marking the its 1$\sigma$ range \citep{watson11}. In upper panel the black solid line shows the {linear regression relation} given in Table \ref{fit}. The symbols and colors have the same meaning as in Fig. \ref{depletions}.}
\label{metals}
\end{center}
\end{figure}

\citet{zafar13} found that the total metal column and extinction in their sample (26 measurements and 21 limits) follows the LG relation ($N_{H,X}=2.2\times10^{21}$cm$^{-2}A_V$ with $N_{H,X}/N$(\hi)$=1.1$) derived by \citet{watson11} from the photoelectric absorption of the soft X-ray of GRBs using the Galactic dust \citep{schlegel98} and \hi\ column density \citep{kalberla05}. The \citet{zafar13} sample followed an average value of metals-to-dust ratio of $10^{21.2}$cm$^{-2}A_V$mag$^{-1}$ and a standard deviation of 0.3\,dex. In Fig. \ref{metals}, upper panel, we plot the relation for our enlarged sample together with the LG soft X-ray relation \citep[green dashed line,][]{watson11}. Throughout the paper, we used Pearson correlation to define a relation together with its coefficient and significance. We used linear regression to derive a best-fit slope and intercept. This is performed in \texttt{IDL} using its routine \texttt{FIT\_AND\_CORRELATE}. In case of fitting data with limits, we used \texttt{IDL} routine \texttt{FITEXY} incorporating upper and lower limits \citep[see][]{kelly07}. In Fig. \ref{metals}, a linear regression relation fit to the data gives a best-fit of $(4.05\pm1.03)\times10^{-22}$, similar to the LG relation. The significance of the correlation is reported in Table \ref{fit}. Table \ref{fit} shows that correlation strengths with measurements only data are quite strong but drop when including limits because of adding more scatter. We also look into S depletion issue and re-fit the data by dropping S-based measurements. The correlation coefficient drops a little to $\rho=+0.47$ but the best-fit value is consistent within 1$\sigma$. This suggests that depletion effect is small for S in our sample.

For the two refractory elements the observed equivalent metal column densities are plotted
in the lower panel of Fig. \ref{metals}. Here Si (triangles) is only 
plotted if Fe is not available. A linear regression fit although provides a similar slope 
with that of the volatile elements (Table \ref{fit}) but has a very
low significance. It is notable that for any given $A_V$ the values
are shifted about 1\,dex towards lower metallicity than the LG, thus
confirming the depletion of those two elements. The fact that the
slope of the relation is not changed is consistent with the
conclusion (Fig. \ref{depletions} above) that the depletions do not
correlate with $A_V$. The shift of about 1\,dex is also consistent
with the average $\delta_{\rm (Fe,~Si)}$ of $\approx -1$.

\subsection{Correlation with metal depletion} \label{metdep}


In Fig. \ref{metals} we could also directly see the effect of the
depletion of refractory elements via grain formation. Where the volatile elements basically
followed the curve for the LG, the refractory elements fell on average
about one dex to the left of that, indicating that the residual amount
of those elements was on average in our sample only about 10\% .

\begin{figure}
\begin{center}
{\includegraphics[width=\columnwidth,clip=]
{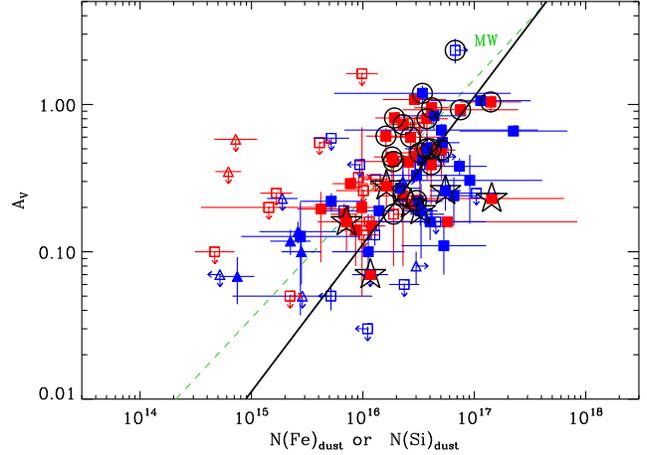}}
\caption{Extinctions as a function of dust-phase column densities. The
green dashed line corresponds to the MW relation given in \citep{vladilo06}.
The black solid line represents the linear regression relation with the best-fit value of $A_V/{\rm N(Fe)}_{\rm dust}=(1.13\pm0.84)\times10^{-17}$. The symbols and colors have the same meaning as in
Fig. \ref{depletions}.  }
\label{nfe}
\end{center}
\end{figure}

Perhaps the best way to test the direct relation between $A_V$ and
metal column density is to consider the elements which are depleted
onto dust, such as Fe, but to add together the residual fraction left
in the gas phase and that which has been depleted onto grains.
\citet{vladilo06} showed that it is possible to compute e.g. the Fe
dust-phase column density (N(Fe)$_{\rm dust}$)
for individual systems using 
\begin{equation}
{\rm N(Fe)}_{\rm dust} = (1-10^{\delta_{\rm Fe}} ){\rm N(Zn})
\bigg( \frac{{\rm Fe}}{{\rm Zn}} \bigg)_\odot
\label{eq:vladiloFe}
\end{equation} 
which is based on the assumption that intrinsically all absorbers
have a Zn to Fe ratio identical to the solar. A similar equation
can be written for S and Si 
\begin{equation}
{\rm N(Si)}_{\rm dust} = (1-10^{\delta_{\rm Si}} ){\rm N(S})
\bigg( \frac{{\rm Si}}{{\rm S}} \bigg)_\odot
\end{equation} 
based on a similar assumption about the intrinsic S to Si ratio. In 
Fig.~\ref{nfe}
we plot $A_V$ vs ${\rm N(Fe)}_{\rm dust}$. Again, as above, in
cases where ${\rm N(Fe)}_{\rm dust}$ is not available, but where we have
${\rm N(Si)}_{\rm dust}$, we plot the latter. In this plot we use the
{\it actual} dust grain column density of the element, i.e. we are not
converting as we did in Fig.~\ref{metals}. This means that we expect a
small horizontal offset between the Fe and Si relation in Fig.~\ref{nfe}, an offset which is much smaller
than the scatter and for the purpose of a visual comparison we can 
ignore it. In the figure we again compare to the DTM relation
of the MW (dashed green line), but here to the relation determined by
\citet{vladilo06} using the same method as defined in equation 
\ref{eq:vladiloFe}.
The agreement with our local environment in the MW is seen to be
very good, the fit parameters are again provided in Table~\ref{fit}.

\begin{figure}
\begin{center}
{\includegraphics[width=\columnwidth,clip=]
{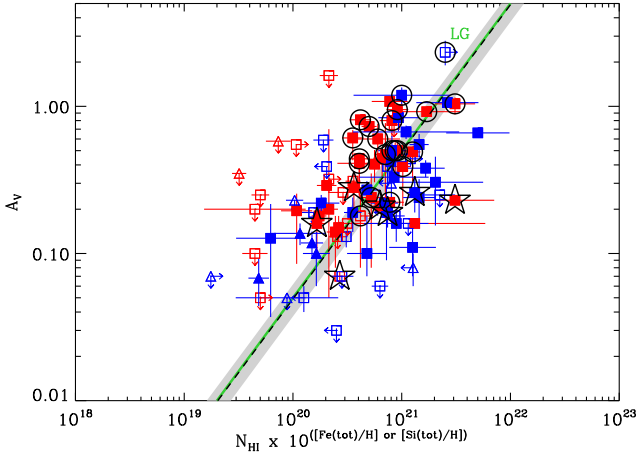}}
\caption{Total column density for the refractory (Fe or Si) elements. The green solid line is the LG relation given in \citet{watson11}. The black dashed line is the linear regression relation where grey shaded region represents error on it. The symbols and colors have the same meaning as in Fig. \ref{depletions}.  }
\label{nfefe}
\end{center}
\end{figure}

With this we are now able to compute the total column density also for the
refractory elements. As noted, the offset in lower panel of Fig. \ref{metals} is due to dust depletion and if we correct for dust column, we would expect to see the similar behaviour of the lower panel as upper panel. We add ${\rm N(Fe)}_{\rm dust}$ and ${\rm N(Si)}_{\rm dust}$ to the observed gas phase column densities in order to obtain the total column densities of refractory elements (since ${\rm N(X)}_{\rm tot} = {\rm N(X)}_{\rm dust} + {\rm N(X)}_{\rm obs}$). We can then derive the equivalent metal column densities of refractory elements (Fig.~\ref{nfefe}) in the same way as we did the gas phase columns in Fig.~\ref{metals} i.e. first converting to equivalent \hi\ column density.

The best-fit values of our and LG sample are in agreement. In fact the agreement is so complementary that we had to increase the linewidth of the LG curve and show our fit as a dashed line, because the fits were covering each other. We also plotted the error on the fit as gay shaded region (see Fig. \ref{nfefe}), however, not that the scatter is not that small. Note that the correlation coefficient strength is not increased significantly when using the total equivalent column densities for refractory elements because of the internal scatter and inclusion of limits in the fit but the significance indicates the reliability of the fit. The agreement of LG relation and depletion-correction derived dust-to-metals best-fit values suggests that the time lag between dust formation after metal formation is small and they both grow together across different environments at all redshifts (see \citealt{zafar13} \S3 for more discussions).

\section{Discussion}

We have assembled the by far largest sample of sightlines (93) for which
both $A_V$ measurements and accurate low ionisation metal column densities
are known. In this sample we have only included sightlines which contain measurements
of at least one of the pairs (Zn, Fe), (S, Si) thereby allowing us always to
correct at least one of the elements (Fe, Si) for assumed depletion.
Starting from the lowest level of assumptions, i.e. simply testing if
individual observed metal column densities scale with $A_V$, and moving to
assuming that deviations from relative solar abundances can be interpreted
as depletion, i.e. as metals removed from the gas-phase by dust grain growth,
we have tested for correlations in the sample.

All the expected correlations are present in our sample. While this does not yet constitute a final confirmation, we conclude that our sample does indeed contain relations which are in full agreement with the conjectures that ($i$) the amount of dust scales directly with the amount of metals available and ($ii$) the dust column density also scales with the amount of dust. Note that there are also nucleosynthetic effects present which could be as large as 0.3\,dex for DLAs \citep{rafelski12,decia16,jenkins17,decia18}. As such this sample should prove a valuable tool to test and refine the current paradigms for dust creation and destruction. In particular its wide redshift coverage ($z=0.4-6.3$) provides support for studies aiming at identifying dust production mechanisms throughout the history of the universe.

\subsection{Intervening non-related reddening}

Along all lines of sight out to high redshift targets, a random number
of intervening unidentified absorbers will be present.  The combined
effect of those absorbers (in case they contain dust) will be an
additional reddening signature, but because the wavelength
scales of dust absorption signatures
are very large, individual random absorbers cannot be identified. Here
we show that we can use our conclusions from the current work to compute
the average statistical signature.

\begin{figure}
\begin{center}
{\includegraphics[width=\columnwidth,clip=]
{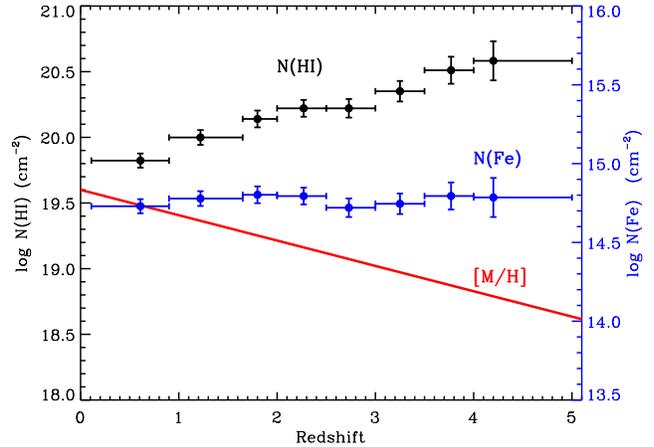}}
\caption{Total N(\hi ) per redshift bin against redshift is shown as black symbols. Total N(\hi ) for each redshift bin is computed using absorber number density, $d(n)/dz$, from \citet{zafar13b}. The average metallicity evolution (offset for illustration) from \citet{rafelski12} is shown as red curve. Blue points indicate total Fe (including both gas and dust-phase) calculated using total N(\hi ) and metallicity.}
\label{nzn}
\end{center}
\end{figure}

We have shown here that there is a strong dependence of $A_V$ on low
ionisation metal column densities. Low ion metals can only exist if they
are shielded from ionising radiation by \hi . The amount of \hi\ along
a random sightline as a function of redshift has been extensively studied 
\citep[e.g.,][]{zwaan05,lah07,noterdaeme12b,zafar13b}. Here we shall use the recent results from \citet{zafar13b} which provides the average number of DLAs per sightline in bins of $\Delta z \sim 0.5$ out to $z=5$. From their Table 6 we compute the average \hi\ column density of DLAs in their sample at each redshift they studied, and find that it is effectively constant with redshift, log N(\hi ) = 20.85. DLAs are
known to make up $\approx 85$\% of all \hi\ \citep{zafar13b}. We correct for
the remaining 15\% (subtracting log(0.85) from the average) obtaining
20.92 which we then multiply on to each redshift bin to obtain the
average N(\hi ) in each bin (shown as black points in Fig. \ref{nzn}).

 \begin{table}
\caption{Pearson linear correlation coefficient, $\rho$ and significance, $\alpha$, together with the best-fit slope as $y/x$.}      
\label{fit} 
\centering     
\setlength{\tabcolsep}{1pt}
\begin{tabular}{l c c c c}  
\hline\hline                        
$x$ & $y$ & $y/x$ & $\rho$ & $\alpha$ \\
\hline
Measurements only & \\
\hline
$\delta_{\rm (Fe, Si)}$ & $A_V$ & $\cdots$ & $-0.05$ & $>$$28\%$ \\ 
N(\hi) $\times10^{{\rm[Zn,S/H]}}$ & $A_V$ & $(3.98\pm1.11)\times10^{-22}$ & $+$0.59 & $>$$99\%$ \\  
N(\hi) $\times10^{{\rm [Fe,Si/H]}}$ & $A_V$ & $\cdots$ & $+$0.06 & $>$$36\%$ \\  
N(Fe, Si)$_{\rm dust}$ & $A_V$ &$(1.12\pm0.78)\times10^{-17}$ & $+$0.62 & $>$$99\%$ \\  
N(\hi)$\times10^{{\rm [Fe_{\rm tot},Si_{\rm tot}/H]}}$ & $A_V$ &$(6.92\pm1.05)\times10^{-22}$ & $+$0.60 & $>$$99\%$ \\  
\hline
\multicolumn{3}{l}{Measurements and limits included} \\
\hline
$\delta_{\rm (Fe, Si)}$ & $A_V$ & $\cdots$ & $-0.16$ & $\approx$$87\%$ \\ 
N(\hi) $\times10^{{\rm [Zn,S/H]}}$ & $A_V$ & $(4.05\pm1.03)\times10^{-22}$ & $+$0.49 & $>$$99\%$ \\  
N(\hi) $\times10^{{\rm[Fe,Si/H]}}$ & $A_V$ & $\cdots$ & $+$0.21 & $\approx$$95\%$ \\  
N(Fe, Si)$_{\rm dust}$ & $A_V$ &$(1.13\pm0.84)\times10^{-17}$ & $+$0.47 & $>$$99\%$ \\  
N(\hi)$\times10^{{\rm [Fe_{\rm tot},Si_{\rm tot}/H]}}$ & $A_V$ &$(4.91\pm0.98)\times10^{-22}$ & $+$0.51 & $>$$99\%$ \\  
\hline
\end{tabular}
\end{table}

The average metallicity evolution of the \hi\ gas has been vividly
debated in the past, but the consensus is now that there is (as one
would also logically expect) a general increase in metallicity as a
function of cosmic time. Here we use the `best fit line' from figure
12 in \citet{rafelski12}. It has a slope of $-0.193$ dex per unit
redshift, and for illustration purposes we plot it in red in Fig.
\ref{nzn} at a convenient offset. Using the average metallicity of each
redshift bin we now compute how much Fe (total, including both gas-phase
and dust-phase) there is in each bin. This is plotted as the blue
points. It is seen that the higher line-of-sight \hi\ column density
per unit redshift in the past is perfectly balanced by the redshift
evolution of metallicity, leaving the amount of metals per unit
redshift remarkably constant (note that the first two and the last bin
are slightly larger than 0.5).  At high redshifts, the change in relative contributions of dust producers \citep{bolmer17,zafar18} could also change the effect of total Fe on extinction.

It is now simple to compute the average total column density of Fe along a
sightline back to a given redshift, and then use the fit in Fig. \ref{nfefe}
to convert this to an average `cosmic dust dimming' CDD($z$). In Fig. \ref{nav},
we provide this quantity, CDD($z$), back to a redshift of $z=5$.

\subsection{Impact on high redshift surveys}
\begin{figure}
\begin{center}
{\includegraphics[width=\columnwidth,clip=]
{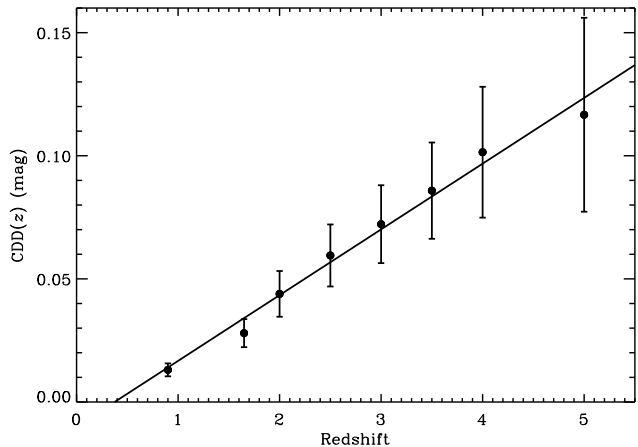}}
\caption{The average `cosmic dust dimming', CDD($z$), is computed as the integrated $A_V$ from redshift 0 to $z$. The $A_V$ is computed from the average amount of dust along the line of sight based on the distribution of Fe (blue points) shown in Fig. \ref{nzn}. }
\label{nav}
\end{center}
\end{figure}

Fig. \ref{nav} shows the accumulated effect of dust in random sightlines out to
cosmological distances. It is seen that up to redshifts of about 3
the CDD($z$) is smaller than the typical uncertainties on individual 
$A_V$ measurements, and the effect can be ignored without too much
impact. At redshifts in excess of 4, in the current paradigm, the CDD($z$) 
continues to rise despite the dropping metallicity, causing the
cumulated effect to become significant. It should be recalled that
the curve shows the average, but the effect is highly stochastic.
Therefore corrections and predictions should probably only be 
considered for statistical samples while the effect on individual
objects will be stochastic.

\subsection{Inclusion of 2175\,\AA\ bump}
The shape of the extinction curve tells us about the dust grains in the ISM \citep{draine03}. Carbonaceous dust grains are responsible for the 2175\,\AA\ bump and silicates induce the steepness in the extinction curve \citep{weingartner01,draine03}. We here aim to see if different extinction curves carrying different dust grain properties could effect our results. Our sample mostly has preference towards featureless Small Magellanic Cloud (SMC)-type extinction curve. However, 21 cases (3 GRBs and 18 QSO-DLAs) are reported to contain a 2175\,\AA\ bump, suggesting carbonaceous dust enriched environments. These cases are highlighted by stars drawn around the data points in Fig. \ref{depletions}--\ref{nfefe}.

\citet{decia13} presented a sample of 18 GRBs (11 measurements and
7 limits) which included a single burst with a 2175\,\AA\ extinction
feature (GRB\,070802). That single burst did not fall on the $A_V$ and N(Fe)$_{\rm dust}$ relation of their limited sample, and \citet{decia13}
suggested that objects with the 2175\,\AA\ bump might follow a
different relation and therefore should not be included.  
Our larger sample includes 19 sightlines with bumps, and is reaching
extinctions 1.8\,mag higher than the \citet{decia13} sample, but
we see no evidence that the 2175\,\AA\ bump cases follow a different
relation. We also find that inclusion of the 2175\,\AA\ bump cases strengthens
the relation especially at higher extinctions. Depending on the amount of metals available to form dust, different grain compositions and sizes could provide a range of optical
extinction laws.

\subsection{Dust-to-gas ratios and molecules}

We would expect a general trend that on average an absorber with a large N(\hi) should
have a large $A_V$.  The general trend will be modified both by the
intrinsic scatter visible in figures \ref{metals}-\ref{nfefe}, but also by the significant
width of the metallicity distribution. In addition to this there are
systematic selection differences between GRBs and QSOs, e.g. that
GRBs are found closer to the centres of galaxies and at higher
redshifts than QSO sightlines, causing their N(\hi) to be systematically
large and metallicities to be systematically low. Such effects will
distort the N(\hi) versus $A_V$ and the N(\hi) versus 
N(Fe,\,Si)$_{\rm dust}$ relations.

Molecular gas is expected to have an even stronger correlation with
dust than the neutral gas. Dust plays a major role for the presence
of molecular gas by catalysing the formation of molecules on the
surface of dust grains as well as shielding against Lyman-Werner
photons. However, our sample contains only seven sightlines with
available N(H$_2$) measurements, and no obvious relation is seen in
the present sample. A larger sample of sightlines with molecule
detections is required for this and should be a primary goal for
future samples.


\section{Conclusions}
In this work, we investigate relations between extinctions, gas and dust-phase properties of the ISM. We use a sample of 46 GRB-DLAs and 47 QSO-DLAs where elemental column densities (of the refractory and volatile elements) together with extinctions are available.  
The sample is the largest sample for such a study and covers a range of 
$z = 0.4-6.3$, $A_V =0.0-2.33$ mag, log N(\hi)$=19.25-22.70$ and
log N(\znii)$= 12.25-14.30$.  
In order to compute depletions we assume, only where required, that
relative element abundances are solar.

We see no direct relation between $A_V$ and depletion. We find the 
extinctions are highly correlated (with a significance $>99$\%) 
with `total equivalent \hi\ column density' for the volatile
elements (Zn, S) and follow the curve for the LG relation.
The fit to the data with a linear regression relation, results in a best-fit of
DTM $(4.05\pm1.03)\times10^{-22}$. Note that \citet{decia16} reported that Zn and S are least depleted but also suffer from depletion. 

The same test performed directly on the observed column densities
of refractory (Fe, Si) elements finds a shift of about 1\,dex from
the LG relation, reflecting the average depletion. Applying the
assumption of relative solar abundances, and using the method defined
by \citet{vladilo06} to calculate the Fe dust-phase column density,
N(Fe)$_{\rm dust}$, we then compute the amount of refractory elements
depleted onto dust grains.  
We find that the refractory element dust-phase column density follows
the MW curve. We add the N(Fe)$_{\rm dust}$ and N(Si)$_{\rm dust}$
to their observed gas-phase column densities to obtain the `total
equivalent \hi\ column densities' for the refractory elements.  
This brings the refractory elements total equivalent metal column densities and extinctions in agreement with the LG DTM relation with a best-fit of $(4.91\pm0.98)\times10^{-22}$. Our methods here providing a way to determine the DTM use direct observables and only the assumption of relative solar abundances.

Last, we show that our results can be used to compute the average 
statistical extinction signature of intervening absorbers. The
amount of \hi\ per unit redshift is known to decrease
with cosmic time while the metallicity of this gas is known to
increase with cosmic time. Remarkably, together those two effects
cause total Fe per unit redshift (both dust and gas-phase) to be almost constant throughout at redshifts
0 to 5.  Integrating the total metal column density (Fe in our case)
out to any redshift, we compute the average `cosmic dust dimming',
CDD($z$), finding a steady increase in dimming (extinction) by
intervening absorbers over increasing distance. The CDD($z$) becomes an
issue that one should consider only at redshifts higher than
about 3, but we caution that the effect is stochastic, and
that correction for individual objects will be highly uncertain.


\begin{table*}
\caption{GRBs data used to estimate dust-to-metals ratios and N(Fe, Si)$_{\rm dust}$. The columns are (1) GRB name, (2) extinction ($A_V$), (3) $z_{\rm abs}$, (4) log N(\hi), (5) log N(\znii), (6) log N(\feii), (7) log N(\sii), (8) log N(\siii), and (9) References to original data.}     
\label{GRB} 
\centering     
\setlength{\tabcolsep}{5pt}
\renewcommand{\footnoterule}{}  
\begin{tabular}{l c c c c c c c c}  
\hline\hline                        
GRB & $A_V$ & $z_{\rm abs}$ & log N(\hi) & log N(\znii) & log N(\feii) & log N(\sii) & log N(\siii) & Refs.\\
 &  mag  & & cm$^{-2}$ & cm$^{-2}$ & cm$^{-2}$ & cm$^{-2}$ & cm$^{-2}$ & \\
\hline
000926 & $0.38\pm0.05$ & 2.038 & $21.30\pm0.20$ & $13.82\pm0.05$ & $15.60\pm0.20$ & $\cdots$ & $16.47\pm0.13$ & 1, 2\\
010222 & $0.24^{+0.08}_{-0.09}$ & 1.475 & $\cdots$ & $13.78\pm0.07$ & $15.32\pm0.15$ & $\cdots$ & $16.09\pm0.05$ & 2, 3 \\
050401 & $0.65\pm0.04$ & 2.899 & $22.60\pm0.30$ & $14.30\pm0.30$ & $16.00\pm0.20$ & $\cdots$ & $16.50\pm0.40$ & 4, 5 \\
050730 & $0.12\pm0.02$ & 3.969 & $22.15\pm0.05$ & $\cdots$ & $\cdots$ & $15.34\pm0.10$ & $15.47\pm0.03$ & 6, 5 \\
050820A & $0.27\pm0.04$ & 2.615 & $21.05\pm0.10$ & $13.28\pm0.04$ & $14.82\pm0.12$ & $15.57\pm0.04$ & $>15.43$ &  7, 8 \\
050922C & $0.07\pm0.02$ & 2.198 & $21.55\pm0.10$ & $\cdots$ & $14.58\pm0.26$ & $14.87\pm0.05$ & $14.97\pm0.07$ & 9,10 \\
051111 & $0.39^{+0.11}_{-0.10}$ & 1.549 & $\cdots$ & $13.47\pm0.04$ & $15.32\pm0.01$ & $\cdots$ & $>16.14$ & 10, 11 \\
060418 & $0.13^{+0.01}_{-0.02}$ & 1.490 & $\cdots$ & $13.09\pm0.01$ & $15.22\pm0.03$ & $\cdots$ & $15.92\pm0.03$ & 12, 10, 11 \\
060714 & $0.21\pm0.02$ & 2.711 & $21.80\pm0.10$ & $13.44\pm0.12$ & $15.01\pm0.05$ & $15.79\pm0.07$ & $15.99\pm0.07$ & 5, 13 \\
061121 & $0.55^{+0.05}_{-0.08}$ & 1.315 & $\cdots$ & $13.76\pm0.06$ & $16.20\pm0.03$ & $\cdots$ & $\cdots$ & 7, 13, 14 \\
070802 & $1.19\pm0.15$ & 2.455 & $21.50\pm0.20$ & $13.60\pm0.60$ & $16.10\pm0.10$ & $\cdots$ & $16.60\pm0.30$ & 15, 5 \\
080210 & $0.33\pm0.03$ & 2.641 & $21.90\pm0.10$ & $13.53\pm0.14$ & $15.98\pm0.37$ & $\cdots$ & $16.20\pm0.13$ & 16, 5  \\
080319C & $0.67\pm0.07$ & 1.949 & $\cdots$ & $13.64\pm0.60$ & $14.94\pm0.14$ & $\cdots$ & $\cdots$ & 13, 14, 17 \\
080413B & $0.84\pm0.16$ & 1.101 & $\cdots$ & $13.57\pm0.15$ & $14.51\pm0.16$ & $\cdots$ & $\cdots$ & 7, 13, 14 \\
080605 & $0.5^{+0.13}_{-0.10}$ & 1.640 & $\cdots$ & $13.53\pm0.08$ & $14.66\pm0.11$ & $\cdots$ & $15.88\pm0.10$ & 18, 13, 14 \\
080905B & $0.42\pm0.03$ & 2.374 & $<22.15$ & $13.52\pm0.13$ & $15.50\pm0.10$ & $\cdots$ & $\cdots$ & 5, 13, 14  \\
081008 & $0.20\pm0.05$ & 1.968 & $21.11\pm0.10$ & $13.15\pm0.04$ & $15.42\pm0.04$ & $\cdots$ & $15.75\pm0.04$ &  19 \\
090323 & $0.1\pm0.04$ & 3.577 & $19.62\pm0.33$ & $<12.7$ & $14.91\pm0.05$ & $15.41\pm0.04$ & $15.46\pm0.13$ & 20, 21 \\
090809 & $0.11\pm0.04$ & 2.737 & $21.40\pm0.08$ & $13.70\pm0.25$ & $15.75\pm0.07$ & $\cdots$ & $16.15\pm0.07$  & 22 \\
100219A & $0.14\pm0.03$ & 4.667 & $21.14\pm0.15$ & $\cdots$ & $14.73\pm0.11$ & $15.25\pm0.15$ & $15.15\pm0.25$ & 23, 24 \\
111008A & $0.12\pm0.04$ & 5.000 & $22.30\pm0.06$ & $13.28\pm0.21$ & $16.05\pm0.05$ & $15.71\pm0.09$ & $>16.04$ & 23, 25 \\
120119A & $1.06\pm0.02$ & 1.729 & $22.44\pm0.12$ & $14.04\pm0.25$ & $15.95\pm0.25$ & $\cdots$ & $16.67\pm0.35$ & 26 \\
120716A & $0.30\pm0.15$ & 2.487 & $21.88\pm0.08$ & $13.91\pm0.32$ & $15.65\pm0.45$ & $\cdots$ & $16.48\pm0.45$ & 26 \\
120815A & $0.19\pm0.04$ & 2.360 & $21.95\pm0.10$ & $13.47\pm0.06$ & $15.29\pm0.05$ & $16.22\pm0.25$ & $16.34\pm0.16$ & 23, 27\\
120909A & $0.16\pm0.04$ & 3.929 & $21.61\pm0.06$ & $13.55\pm0.32$ & $15.20\pm0.18$ & $\cdots$ & $16.22\pm0.32$ & 26 \\ 
121024A & $0.26\pm0.06$ & 2.300 & $21.88\pm0.10$ & $13.74\pm0.03$ & $15.82\pm0.05$ & $>15.9$ & $>16.35$ & 23, 28 \\
130408A & $0.22^{+0.04}_{-0.05}$ & 3.758 & $21.76\pm0.03$ & $12.87\pm0.16$ & $15.52\pm0.11$ & $15.78\pm0.18$ & $15.95\pm0.22$ & 27 \\
141028A & $0.13\pm0.09$ & 2.333 & $20.55\pm0.07$ & $12.38\pm0.33$ & $14.23\pm0.21$ & $\cdots$ & $14.82\pm0.33$ &  26, 29 \\
\hline
990123 & $<0.25$ & 1.600 & $\cdots$ & $13.95\pm0.05$ & $14.78\pm0.15$ & $\cdots$ & $\cdots$ & 2, 3 \\
020813 & $<0.18$ & 1.255 & $\cdots$ & $13.54\pm0.06$ & $15.48\pm0.04$ & $\cdots$ & $16.29\pm0.04$ & 30 \\
030226 & $0.05\pm0.01$ & 1.987 & $20.50\pm	0.30$ & $<12.70$ & $14.86\pm0.01$ & $>13.30$ & $15.07\pm0.04$ & 31, 21 \\
030323 & $<0.16$ & 3.371 & $21.90\pm0.07$ & $13.66\pm0.05$ & $15.93\pm0.08$ & $15.84\pm0.19$ & $15.98\pm0.04$ & 32 \\
050505 & $0.30\pm0.10$ & 4.275 & $22.05\pm0.10$ & $\cdots$ & $>15.50$ & $>16.10$ & $>15.70$ & 33, 34 \\
050904 & $<0.05$ & 6.295 & $21.62\pm0.02$ & $\cdots$  & $\cdots$ & $15.14\pm0.17$ & $14.29\pm0.57$ & 24, 35\\
060206 & $<0.23$ & 4.048 & $20.85\pm0.10$ & $\cdots$ & $>14.65$ & $15.21\pm0.05$ & $15.23\pm0.04$ & 7, 36\\
060526 & $<0.39$ & 3.221 & $20.00\pm0.15$ & $<12.91$ & $14.28\pm0.24$ & $14.58\pm0.25$ & $15.87\pm0.16$ & 21, 37 \\
060707 & $0.08\pm0.02$ & 3.425 & $21.00\pm0.20$ & $\cdots$  & $\cdots$  & $>16.30$ & $16.15\pm0.18$ & 5, 38 \\
070110 & $<0.10$ & 2.3521 & $21.70\pm0.10$ & $13.53\pm0.08$ & $14.75\pm0.05$ & $\cdots$ & $15.84\pm0.09$ & 5, 13 \\
070506 & $0.44\pm0.05$ & 2.308 & $22.00\pm	0.30$  & $>13.68$ & $15.50\pm0.20$ & $\cdots$  & $\cdots$ & 5, 13, 14  \\
071031 & $<0.07$ & 2.692 & $22.15\pm0.05$ & $13.05\pm0.03$ & $15.20\pm0.1$ & $\cdots$  & $\cdots$ & 5, 8 \\
080330 & $<0.19$ & 1.511 & $\cdots$  & $12.79\pm0.06$ & $14.70\pm0.09$ & $\cdots$ & $>15.41$ & 39, 40 \\
080413A & $<0.59$ & 2.433 & $21.85\pm0.15$ & $12.88\pm0.07$ & $15.57\pm0.04$ & $\cdots$  & $\cdots$ & 8, 21 \\
080607 & $2.33^{+0.43}_{-0.46}$ & 3.037 & $22.70\pm0.15$ & $>14.00$ & $>16.70$ & $	>16.34$ & $\cdots$ & 5, 41 \\
080721 & $<0.34$& 2.5914 & $21.60\pm0.10$ & $13.69\pm0.05$ & $15.27\pm0.04$ & $\cdots$ & $\cdots$ & 13, 42 \\
090926A & $<0.03$ &  2.107 & $21.60\pm0.07$ & $<13.00$ & $14.86\pm0.09$ & $14.89\pm0.10$ & $14.80\pm0.08$ & 43\\
110205A & $0.19\pm0.10 $ & 2.215 & $21.45\pm0.23$ & $<13.46$ & $14.81\pm0.08$ & $<15.86$ & $14.56\pm0.07$ & 44, 45 \\
120327A & $<0.06$ & 2.815 & $22.01\pm0.09$ & $13.40\pm0.04$ & $15.78\pm0.02$ & $	15.74\pm0.02$ & $16.36\pm0.03$ & 46 \\
130606A & $<0.07$ & 5.913& $19.91\pm0.02$ & $\cdots$ & $13.29\pm0.07$ & $<14.44$ & $13.95\pm0.11$ & 23, 47 \\
\hline
\end{tabular}
\begin{minipage}{180mm}
1: \citep{chen07}, 2: \citep{starling07}, 3: \citep{savaglio03}, 4: \citep{watson06}, 5: \citep{zafar11}, 6: \citep{chen05}, 7: \citep{schady12}, 8: \citep{ledoux09}, 9: \citep{piranomonte08}, 10: \citep{schady10}, 11: \citep{prochaska07}, 12: \citep{vreeswijk07}, 13: this work$^a$, 14: \citep{zafar13}, 15: \citep{ardis}, 16: \citep{decia11}, 17: \citep{perley09}, 18: \citep{zafar12}, 19: \citep{delia11}, 20: \citep{savaglio12}, 21: \citep{schady11}, 22: \citep{skuladottir10}, 23: \citep{zafar18a}, 24: \citep{thoene12}, 25: \citep{sparre14}, 26: \citep{wiseman17}, 27: \citep{kruhler13}, 28: \citep{friis15}, 29:\citep{zafar18}, 30: \citep{savaglio04}, 31: \citep{shin06}, 32: \citep{vreeswijk04}, 33: \citep{berger06}, 34: \citep{hurkett06}, 35: \citep{zafar10}, 36: \citep{thoene08}, 37: \citep{thoene10}, 38: \citep{laskar11}, 39: \citep{delia09a}, 40: \citep{greiner11}, 41: \citep{prochaska09}, 42: \citep{covino13}, 43: \citep{delia10}, 44: \citep{cucchiara11}, 45: \citep{gendre12}, 46: \citep{delia14}, 47: \citep{hartoog15} \\
$^a$ Based on the equivalent width measurements in \citet{fynbo09}
following the procedure described in \citet{laskar11,zafar13}.
 
\end{minipage}
\end{table*}

\begin{table*}
\caption{QSO DLAs and sub-DLAs data used to estimate dust-to-metals ratios and N(Fe, Si)$_{\rm dust}$. The columns are provided similar to as in Table \ref{GRB}.}      
\label{QSO} 
\centering     
\setlength{\tabcolsep}{5pt}
\renewcommand{\footnoterule}{}  
\begin{tabular}{l c c c c c c c c}  
\hline\hline                        
QSO & $A_V$ & $z_{\rm abs}$ & log N(\hi) & log N(\znii) & log N(\feii) & log N(\sii) & log N(\siii) & Refs.\\
 &  mag  & & cm$^{-2}$ & cm$^{-2}$ & cm$^{-2}$ & cm$^{-2}$ & cm$^{-2}$ & \\
\hline
Q\,0000$+$0048 & $0.23\pm0.01$ & 2.526 & $20.80\pm0.10$ & $14.09\pm0.45$ & $15.14\pm0.03$ & $\cdots$ & $15.93\pm0.17$ & 1 \\
Q\,0016$+$0012 & $0.16^{+0.04}_{-0.06}$ & 1.973 & $20.83\pm0.05$ & $12.82\pm0.04$ & $14.81\pm0.03$ & $15.28\pm0.02$  & $15.43\pm0.03$ & 2, 3 \\
Q\,0111$+$0641 & $0.22\pm0.01$ & 2.027 & $21.50\pm0.30$ & $13.50\pm0.10$ & $15.80\pm0.10$ & $\cdots$ & $<16.70$ & 4 \\
Q\,0121$+$0027 & $0.47\pm0.30$ & 1.395 & $\cdots$ & $13.45\pm0.09$ & $14.82\pm0.05$ & $\cdots$ & $15.59\pm0.13$ & 5 \\
Q\,0745$+$4554 & $0.95^{+0.09}_{-0.18}$ & 1.861 & $\cdots$ & $13.56\pm0.03$ & $15.09\pm0.09$ & $\cdots$ & $15.77\pm0.04$ & 6, 7 \\
Q\,0843$+$0221 & $\approx0.07$ & 2.786 & $21.99\pm0.08$ & $13.03\pm0.01$ & $14.96\pm0.01$ & $15.57\pm0.01$ & $15.77\pm0.01$ & 8 \\
Q\,0918$+$1636 & $\approx0.21$ & 2.583 & $20.96\pm0.05$ & $13.40\pm0.01$ & $15.43\pm0.01$ & $15.82\pm0.01$ & $16.01\pm0.01$ & 9 \\
Q\,0927$+$1543 & $0.60\pm0.16$ & 1.731 & $21.35\pm0.15$ & $13.38\pm0.05$ & $15.14\pm0.24$ & $\cdots$ & $15.99\pm0.05$ & 10, 11 \\
Q\,1006$+$1538 & $0.41_{-0.17}^{+0.07}$ & 2.206 & $20.00\pm0.15$ & $13.21\pm0.21$ & $14.36\pm0.12$ & $\cdots$ & $15.30\pm0.19$ & 6, 7  \\
Q\,1007$+$2853  & $1.08\pm0.09$ & $0.884$ & $\cdots$ & $13.49\pm0.17$ & $15.86\pm0.09$ & $\cdots$ & $16.39\pm0.20$ & 12 \\
Q\,1047$+$3423 & $0.61_{-0.18}^{+0.08}$ & 1.669 & $20.05\pm0.20$ & $13.15\pm0.09$ & $14.69\pm0.02$ & $\cdots$ & $15.82\pm0.13$ & 6, 7 \\
Q\,1130$+$1850 & $0.49_{-0.18}^{+0.04}$ & 2.012 & $21.10\pm0.30$ & $13.70\pm0.03$ & $15.91\pm0.01$ & $\cdots$ & $16.60\pm0.02$ & 6, 7 \\
Q\,1141$+$4442 & $0.24_{-0.16}^{+0.11}$ & 1.902 & $20.85	\pm0.15$ & $13.32\pm0.12$ & $15.22\pm0.03$ & $\cdots$ & $\cdots$ & 6 \\
Q\,1157$+$6135 & $0.92\pm0.07$ & 2.459 & $21.80\pm0.20$ & $13.83\pm0.33$ & $15.63\pm0.19$ & $\cdots$ & $16.48\pm0.48$ & 13 \\
Q\,1157$+$6155 & $1.04_{-0.16}^{+0.11}$ & 2.460 & $21.80\pm0.20$ & $14.09\pm0.11$ & $15.63\pm0.19$ & $\cdots$ & $\cdots$ & 6 \\
Q\,1159$+$0112 & $0.14^{+0.04}_{-0.06}$ & 1.944 & $21.60\pm0.10$ & $12.99\pm0.05$ & $15.46\pm0.02$ & $\cdots$ & $15.97\pm0.01$ & 14, 3\\
Q\,1211$+$0833 & $0.80^{+0.05}_{-0.18}$ & 2.117 & $21.00\pm0.20$ & $13.51\pm0.09$ & $14.70\pm0.07$ & $\cdots$ & $15.65\pm0.07$ & 15 \\
Q\,1237$+$0647 & $0.15\pm0.03$ & 2.690 & $20.00\pm0.15$ & $13.02\pm0.02$ & $14.57\pm0.01$ & $15.39\pm0.06$ & $15.15\pm0.02$ & 16 \\
Q\,1321$+$2135 & $0.39_{-0.18}^{+0.03}$ & 2.125 & $21.55\pm0.20$ & $13.61\pm0.03$ & $15.84\pm0.02$ & $\cdots$ & $16.26\pm0.02$ & 6, 7 \\
Q\,1323$-$0021 & $0.44^{+0.08}_{-0.11}$ & $0.716$ & $20.21\pm0.20$ & $13.43\pm0.05$ & $15.15\pm0.03$ & $\cdots$ & $\cdots$ & 17, 3 \\ 
Q\,1422$-$0001 & $0.29\pm0.01$ & 0.909 & $20.40\pm0.40$ & $12.91\pm0.07$ & $15.26\pm0.03$ & $\cdots$ & $15.57\pm0.07$ & 18 \\
Q\,1439$+$1117 & $0.20^{+0.50}_{-0.15}$ & 2.419 & $20.10\pm0.10$ & $12.93\pm0.04$ & $14.28\pm0.05$ & $15.27\pm0.06$ & $14.80\pm0.04$ & 19, 20 \\
Q\,1459$+$0024  & $0.81\pm0.03$ & 1.389 & $\cdots$ & $13.22\pm0.08$ & $14.46\pm0.05$ & $\cdots$ & $<15.46$ & 5 \\
Q\,1524$+$1030 & $0.50_{-0.18}^{+0.07}$ & 1.940 & $21.45\pm0.10$ & $13.56\pm0.01$ & $15.26\pm0.07$ & $\cdots$ & $\cdots$ & 6 \\
Q\,1531$+$2403 & $0.44_{-0.19}^{+0.03}$ & 2.002 & $20.20\pm0.20$ & $13.21\pm0.06$ & $14.90\pm0.06$ & $\cdots$ & $15.82\pm0.01$ & 6, 7 \\
Q\,1604+2203 & $0.73\pm0.05$ & 1.640 & $\cdots$ & $13.30\pm0.02$ & $14.67\pm0.01$ & $\cdots$ & $15.13\pm0.03$ & 21 \\
Q\,1705$+$3543 & $0.40\pm0.10$ & 2.038 & $20.62\pm0.12$ & $13.35\pm0.07$ & $14.64\pm0.04$ & $\cdots$ & $15.29\pm0.14$ & 22 \\
Q\,1737$+$4406 & $0.50^{+0.07}_{-0.19}$ & 1.614 & $\cdots$ & $13.54\pm0.04$ & $15.48\pm0.07$ & $\cdots$ & $15.89\pm0.06$ & 6, 7 \\
Q\,2140$-$0321 & $0.16\pm0.01$ & 	2.340 & $22.40\pm0.10$ & $13.72\pm0.92$ & $15.64\pm0.03$ & $\cdots$ & $\cdots$ & 23, 24\\	
Q\,2222$-$0946 & $0.18\pm0.03$ & 2.350 & $20.65\pm0.05$ & $12.83\pm0.01$ & $15.13\pm0.01$ & $15.31\pm0.01$ & $15.62\pm0.01$ & 25 \\
Q\,2225$-$0527 & $0.28\pm0.08$ & 2.130 & $20.69\pm0.05$ & $13.16\pm0.02$ & $14.87\pm0.01$ & $15.41\pm0.02$ & $15.49\pm0.04$ & 26 \\
Q\,2340$-$0052 & $0.21^{+0.06}_{-0.11}$ & 1.360 & $\cdots$ & $12.62\pm0.05$ & $14.99\pm0.04$ & $\cdots$ & $15.70\pm0.02$ & 3 \\
\hline
Q\,0013$+$0004 & $ <0.10$ & 2.025& $20.80\pm0.10$ & $12.25\pm0.05$ & $15.21\pm0.11$ & $14.91\pm0.04$ & $15.31\pm0.04$ & 3 \\
Q\,0816$+$1446 & $<0.75$ & 3.287 & $22.00\pm0.10$ & $13.53\pm0.01$ & $15.89\pm0.02	$ & $\cdots$ & $16.31\pm0.01$ & 27 \\
Q\,0938$+$4128 & $<0.20$ & 1.373 & $20.52\pm0.10$ & $12.25\pm0.10$ & $14.82\pm0.58$ & $\cdots$ & $\cdots$ & 3 \\	     	    
Q\,0948$+$433 & $<0.31$ & 1.233 & $21.62\pm0.06$ & $13.15\pm0.01$ & $15.56\pm0.01$ & $\cdots$ & $>15.56$ & 3 \\    	     		
Q\,1010$+$0003  & $<0.13$ & 1.265 & $21.52\pm0.07$ & $13.01\pm0.02$ & $15.26\pm0.05$ & $\cdots$ & $\cdots$ & 3 \\	  
Q\,1107$+$0048  & $<0.26$ & 0.741 & $21.0	0\pm0.05$ & $13.06\pm0.15$ & $15.53\pm0.02$ & $\cdots$ & $\cdots$ & 3 \\
Q\,1209$+$6717 & $0.18^{+0.07}_{-0.10}$ & 1.843 & $20.25\pm0.20$ & $<13.22$ & $14.83\pm0.01$ & $\cdots$ & $\cdots$ & 6 \\
Q\,1232$-$0224 & $<0.32$ & 0.395 & $20.75\pm0.07$ & $12.93\pm0.12$ & $<14.94$ & $\cdots$ & $\cdots$ & 3 \\
Q\,1439$+$1117 & $<1.62$ & 2.418 & $20.10\pm0.10$ & $12.93\pm0.04$ & $14.28\pm0.05$ & $15.27\pm0.06$ & $14.80\pm0.04$ & 10, 19\\	
Q\,1501$+$0019 & $<0.16$ & 1.483 & $20.85\pm0.05$ & $13.10\pm0.05$ & $15.53\pm0.03$ & $\cdots$ & $\cdots$ & 3 \\	
Q\,2123$-$0050 & $<0.35$ & 2.060 & $19.25\pm0.20$ & $\cdots$ & $14.12\pm0.02$ & $14.70\pm0.02$ & $14.69\pm0.02$ & 10, 28 \\	
Q\,2234$+$0000  & $<0.25$ & 2.066 & $20.59\pm0.08$ & $12.30\pm0.05$ & $14.83\pm0.03$ & $15.15\pm0.02$ & $15.39\pm0.06$ & 3, 29 \\
Q\,2239$-$2949 & $<0.05$ & 1.825 & $	19.84\pm0.14$ & $<12.30$ & $14.11\pm0.04$ & $\cdots$ & $14.68\pm0.06$ & 30 \\
Q\,2340$-$0053 & $<0.55$ & 2.054 & $20.33\pm0.03$ & $12.63\pm0.08$ & $>14.97$ & $\cdots$ & $15.17\pm0.04$ & 10, 31 \\ 
Q\,2350$-$0052 & $<0.58$ & 2.426 & $20.50\pm0.10$ & $<12.20$ & $14.83	\pm0.07$ & $15.06\pm0.10$ & $15.26\pm0.07$ & 10, 32 \\ 
\hline
\end{tabular}
\begin{minipage}{180mm}
1: \citep{noterdaeme17}, 2: \citep{petitjean02}, 3: \citep{vladilo06}, 4: \citep{fynbo17}, 5: \citep{jiang10}, 6: \citep{ma18}, 7: \citep{ma17}, 8: \citep{balashev17}, 8: \citep{fynbo11}, 10: \citep{ledoux15}, 11: \citep{berg15}, 12: \citep{zhou10}, 13: \citep{wang12}, 14: \citep{dessauges07}, 15: \citep{ma15}, 16: \citep{noterdaeme10}, 17: \citep{peroux06}, 18: \citep{bouche16}, 19: \citep{noterdaeme08a}, 20: \citep{rudie17}, 21: \citep{noterdaeme09b}, 22: \citep{pan17}, 23: \citep{noterdaeme15}, 24: \citep{noterdaeme15b}, 25: \citep{krogager13}, 26: \citep{krogager16}, 27: \citep{guimaraes12}, 28: \citep{milutinovic10}, 29: \citep{vladilo11}, 30: \citep{zafar17}, 31: \citep{prochaska07}, 32: \citep{Dessauges03}
\end{minipage}
\end{table*}


\section*{Acknowledgements}


\bibliographystyle{aa}
\bibliography{depletions.bib}{}

\bsp

\label{lastpage}
\end{document}